\documentclass[11pt,fleqn]{article}

\usepackage{amssymb}

\newcommand{\bls}[1]{\renewcommand{\baselinestretch}{#1}}

\def\noi{\noindent}

\newcommand{\Title}[1]{\noi {{\Large\bf #1}}\\[1ex]}

\newcommand{\Author}[2]{\noi{\bf #1}\\[2ex]\noi{\small\it #2}\\}

\newcommand{\Abstract}[1]{\vskip 2mm \begin{center}
        \parbox{16.4cm}{\small\noi #1} \end{center}\medskip}

\newcommand{\foom}[1]{\protect\footnotemark[#1]}

\def\email#1#2{\footnotetext[#1]{e-mail: #2}\addtocounter{footnote}{1}}

\def\nqq{\hspace*{-2em}}

\def\cm{\hspace*{1cm}}
\def\inch{\hspace*{1in}}

\def\Jl#1#2{#1 {\bf #2},\ }

\def\ApJ#1 {\Jl{Astroph. J.}{#1}}
\def\CQG#1 {\Jl{Class. Quantum Grav.}{#1}}
\def\DAN#1 {\Jl{Dokl. AN SSSR}{#1}}
\def\GC#1 {\Jl{Grav. Cosmol.}{#1}}
\def\GRG#1 {\Jl{Gen. Rel. Grav.}{#1}}
\def\JETF#1 {\Jl{Zh. Eksp. Teor. Fiz.}{#1}}
\def\JETP#1 {\Jl{Sov. Phys. JETP}{#1}}
\def\JHEP#1 {\Jl{JHEP}{#1}}
\def\JMP#1 {\Jl{J. Math. Phys.}{#1}}
\def\NPB#1 {\Jl{Nucl. Phys. B}{#1}}
\def\NP#1 {\Jl{Nucl. Phys.}{#1}}
\def\PLA#1 {\Jl{Phys. Lett. A}{#1}}
\def\PLB#1 {\Jl{Phys. Lett. B}{#1}}
\def\PRD#1 {\Jl{Phys. Rev. D}{#1}}
\def\PRL#1 {\Jl{Phys. Rev. Lett.}{#1}}

\def\lal{&&\nqq {}}
\def\eq{Eq.\,}
\def\eqs{Eqs.\,}
\def\beq{\begin{equation}}
\def\eeq{\end{equation}}
\def\bear{\begin{eqnarray}}
\def\bearr{\begin{eqnarray} \lal}
\def\ear{\end{eqnarray}}
\def\earn{\nonumber \end{eqnarray}}
\def\nn{\nonumber\\ {}}

\def\nnn{\nonumber\\ \lal }

\def\yy{\\[5pt] {}}

\def\d{\partial}

\def\diag{\mathop{\rm diag}\nolimits}

\def\const{{\rm const}}
\def\eps{\varepsilon}

\bls{0.98}
\begin{document}
\twocolumn[

\Title{Spin-spin interaction in general relativity  \yy
       and induced geometries with nontrivial topology}

\Author{V. G. Krechet and D. V. Sadovnikov\foom 1}
       {Yaroslavl State Pedagogical University}

\Abstract {We consider the dynamics of a self-gravitating spinor
 field and a self-gravitating rotating perfect fluid. It is shown that both
 these matter distributions can induce a vortex field described by the curl
 4-vector of a tetrad:
   $\omega^i=\frac12\eps^{iklm}e_{(a)k}e^{(a)}_{\ \ l;m}$, where
 $e^{(a)}_k$ are components of the tetrad. The energy-momentum tensor
 $T_{ik}(\omega)$ of this field has been found and shown to violate the
 strong and weak energy conditions which leads to possible formation of
 geometries with nontrivial topology like wormholes. The corresponding exact
 solutions to the equations of general relativity have been found. It is also
 shown that other vortex fields, e.g., the magnetic field, can also possess
 such properties.}

] 
\email 1 {sdv69@mail.ru}

 {
\def\ae{\varkappa}

  As we have shown earlier [1, 2], a self-gravitating Dirac spinor field
  with the Lagrangian
\beq                \label{krec-L(spin)}
    L(\psi)=\frac{\hbar c}{2}
    \left[\nabla_i\bar\psi\gamma^i\psi-\bar\psi\gamma^i\nabla_i\psi-
        F(\bar\psi\psi)\right]
\eeq
  can interact with the vortex component of the gravitational field, which
  results in the appearance of a more general Lagrangian:
\bearr              \label{krec-L2(spin)}
    L(\psi)=\frac{\hbar c}{2}
    \Big[\d_i\bar\psi\gamma^i\psi-\bar\psi\gamma^i\nabla_i\psi
\nnn \cm
    + \omega^i\cdot(\bar\psi\gamma_i\gamma_5\psi)-F(\bar\psi\psi)\Big].
\ear
  Here, $\omega^i$ is the curl 4-vector of a tetrad:
  $\omega^i = \frac12 \eps^{iklm}e^{(a)}_ke_{(a)l;m}$, i.e., the 4-vector
  of the gravitational field vortex; $\nabla_\alpha\psi$ is the covariant
  derivative of the spinor function $\psi(x^k)$:
  $\nabla_k\psi=\d_k\psi-\Gamma_k\psi$, where $\Gamma_k$ are the matrix
  spinor connection coefficients; $\gamma_k$ are the curved-space Dirac
  matrices defined by the fundamental relation between the space-time metric
  and spin,
\[
    \gamma_i\gamma_k+\gamma_k\gamma_i=2g_{ik}\cdot I,
\]
  and the axial vector $\bar\psi\gamma_i\gamma_5\psi$ is proportional to
  the proper angular momentum (spin) of the spinor field
  $S_k(\psi)=\frac{\hbar c}{2}\bar\psi\gamma_k\gamma_5\psi$;
  the function $F(\bar\psi\psi)$ is the spinor field potential depending on
  the invariant $\bar\psi\psi$. In particular, for a massive spinor field
  we have $F(\bar\psi\psi) = 2m\bar\psi\psi$.

  Variation of the total Lagrangian of the gravitational and spinor fields
  $L = - R/(2\ae) + L(\psi)$ with respect to $\omega^i$ leads to a
  relation between the gravitational field vortex and the spin density
  of the spinor field:
\beq        \label{krec-svyaz}
    \omega^i = \ae\frac{\hbar c}{4}\bar\psi\gamma^i\gamma_5\psi.
\eeq
  Taking into account this relation, the spinor field Lagrangian
 (\ref{krec-L2(spin)}) takes the form
\bearr\label{krec-L3(spin)}
    L(\psi) = \frac{\hbar c}{2}
    \Big[\d_k\bar\psi\gamma^k\psi-\bar\psi\gamma^k\d_k\psi
\nnn \cm
    + \frac{\ae hc}{2}
    (\bar\psi\gamma^k\gamma_5\psi)(\bar\psi\gamma_k\gamma_5\psi)
        -F(\bar\psi\psi)\Big],
\ear
  i.e., we have obtained the Lagrangian of a nonlinear spinor field with a
  quadratic pseudovector nonlinearity.

  Interaction of such a nonlinear spinor field with gravity, even if the
  latter has no vortex component, e.g., in the case of spherical symmetry,
  leads to an interesting result. Spherically symmetric spinor field
  configurations have a radially polarized spin density vector
\[
    S_i(\psi) = \frac{\hbar c}{2}\bar\psi\gamma_i\gamma_5\psi
    =\delta_i^1 \bar\psi\gamma_1\gamma_5 \psi\frac{\hbar c}{2},
\]
  distributed like the lines of force of a point electric charge. Let us
  choose the metric of a static, spherically symmetric space-time in the form
\bearr\label{krec-sf-metrica}
    ds^2 = e^{\nu(r)}dt^2-e^{\lambda(r)}dr^2
        -r^2(d\theta^2+\sin^2\theta\,d\varphi^2).
\nnn
\ear
  The components of the energy-momentum tensor of the nonlinear spinor field
  (\ref{krec-L3(spin)}) in the absence of the potential $F(\bar\psi\psi)$
  take the form
\bearr                  \label{krec-tenzor_energy-ipmuls}
    T_{ik}(\psi)
\nnn
    =\frac{hc}{4}\left[\nabla_i\bar\psi\gamma_k\psi+
    \nabla_k\bar\psi\gamma_i\psi-\bar\psi\gamma_i\nabla_k\psi-
        \bar\psi\gamma_k\nabla_i\psi\right]
\nnn \cm
    - \frac{hc}{2}\frac{\ae hc}{2}
    (\bar\psi\gamma^s\gamma_s\psi)(\bar\psi\gamma_s\gamma_5\psi)g_{ik}.
\ear
  Solving the set of Einstein-spinor equations due to the Lagrangian
  (\ref{krec-L3(spin)}) with (\ref{krec-tenzor_energy-ipmuls}),
\bear
    R_{ik}-\frac12 R g_{ik} = \ae T_{ik}(\psi),
\nn
    \gamma^k\nabla_k\psi-\frac{\ae hc}{2}
    (\bar\psi\gamma^k\gamma_5\psi)\gamma_k\gamma_5\psi = 0\
\ear
  in a space-time with the metric (\ref{krec-sf-metrica}), we find the
  function $\psi(r)$ and the metric coefficients $e^{\lambda(r)}$ and
  $e^{\nu(r)}$ [3, 4]. In particular, for the coefficients $e^{\lambda(r)}$
  and $e^{\nu(r)}$ we obtain the expressions $e^\nu =1$ and
  $e^\lambda = {r^2}/(r^2-a^2)$ ($a=\const$), and to keep the signature
  unchanged we must put $r^2-a^2>0$. Then, after the transformation
  $r^2-a^2 = x^2$ ($-\infty<x<+\infty$), the metric (\ref{krec-sf-metrica})
  is obtained in the form
\beq\label{krec-sf-metrica-itog}
    ds^2 = dt^2-dx^2-(x^2+a^2)(d\theta^2+\sin^2\theta\,d\varphi^2).
\eeq

  It is the metric of a wormhole space-time (the so-called Ellis wormhole),
  connecting two asymptotically flat spaces. The constant $a$ determines
  the wormhole throat radius. In this case, $a = l_0 s_0\sqrt{\ae/\eps_0}$,
  where $l_0 =\sqrt{\ae hc}$ is the Planck length while $s_0$ and $\eps_0$
  are the values of the spin flux density and energy density of the spinor
  field at $x=0$. It can be seen that the throat radius of a wormhole formed
  by a polarized self-gravitating nonlinear spinor field is of the order of
  Planck's length. This is connected with the fact that the coefficient
  of nonlinearity $\ae hc/2 = l_0^2$ is of the order of Planck's length
  squared. If, however, this coefficient of nonlinearity can be much greater,
  the wormhole throat will also be much wider.

  The metric (8) was discussed in detail by Ellis [5] and is a special case
  of metrics discussed in [6] in the context of scalar-tensor theories of
  gravity.

  A simple example of a space-time with a stationary vortex gravitational
  field is the space-time with the cylindrically symmetric metric
\bearr\label{krec-tsilindr-metrica}
    ds^2 = D(x)\,dt^2 - A(x)\,dx^2 - B(x)\,d\alpha^2
\nnn \cm
    -A(x)\,dz^2 - 2E(x)\,d\alpha\, dt.
\ear
  The geometric properties of its spatial section are determined by the
  3-dimensional spatial line element
\beq\label{krec-3-element}
    dl^2=A\,dx^2+\frac{BD+E^2}{D}\,d\alpha^2+A\,dz^2,
\eeq
  while the intensity of the stationary gravitational vortex
  $\omega=(\omega_k\omega^k)^{1/2}$ is determined by the expression
\beq\label{krec-intensiv}
    \omega = \frac{E'D - D'E}{2DA^{1/2}(E^2 + BD)^{1/2}},
\eeq
  where the prime denotes $d/dx$. In the spatial metric
  (\ref{krec-3-element}), the coefficient $R(x)\equiv (BD + E^2)/D$ of the
  angular coordinate squared determines the length of the coordinate
  circle $x=x_0$, $z=z_0$. From the vacuum Einstein equations $R_{ik} = 0$
  for the metric (\ref{krec-tsilindr-metrica}) one can obtain an equation
  for the coefficient $R(x)$:
\beq\label{krec-R(x)}
    A^{-1}\left[\frac{R''}{R}+\frac{R'}{2R}\left(\frac{D'}{D}-
    \frac{R'}{R}\right)\right]=\frac{4\omega^2}{c^2}.
\eeq
  The right-hand side in \eq (\ref{krec-R(x)}), proportional to
  $\omega^2$, is positive-definite. Therefore, at the point where $R'=0$,
  one obtains $R''>0$ (a minimum), and this is a necessary condition for
  the existence of a wormhole. Thus a vortex gravitational field is able
  to induce wormhole formation. The definitions and general conditions for
  the existence of static cylindrical wormholes have been considered in [7].

  A solution to the vacuum Einstein equations $R_{ik}=0$ for the metric
  under consideration describes a wormhole space-time:
\bearr\label{krec-Reshenie_R_ik=0}
    A(x) = \frac{c}{b\omega_0(x^2/b^2 +1)^2};
\nnn
    D(x) = \frac{\exp\left[\arcsin (x^2/b^2 +1)^{-1}\right]}
        {x^2/b^2 +1};
\nnn
    R(x) = \frac{BD+E^2}{D}
\nnn \ \ \
    =(x^2+b^2) \exp\left[\arcsin (x^2/b^2 +1)^{-1}\right];
\nnn
    \omega = \frac{\omega_0}{AD^{1/2}};\qquad
    (\omega_0=\const,\ b=\const),
\ear
  where $-\infty < x < \infty$. It is seen here that the ``circular''
  metric coefficient $R(x)$, determining the coordinate circumference,
  nowhere turns to zero, and at the point $x=0$ (throat) has a minimum:
  $R(x)_{\min} = b^2\cdot \exp(\pi/2)\ne 0$, so that the throat radius
  of the obtained wormhole is $a=b\cdot \exp(\pi/4)$. Here $b$ is an
  integration constant. Thus a free vortex gravitational field can form
  cylindrical wormholes, space-time tunnels connecting different regions of
  space-time.

  Wormholes can be also induced by vortex fields other than spinor and
  gravitational ones, for example, an azimuthal magnetic field $H_\alpha$.
  (A solution to the Einstein-Maxwell equations for different directions of
  electric and magnetic fields was obtained earlier in [8].)

  In a static, cylindrically symmetric space-time with an azimuthal
  magnetic field n$H_\alpha=F_{13}$, described by the metric
\bearr\label{krec-tsilindr-metrica2}
    ds^2 = D(x)\,dt^2 - A(x)\,dx^2
\nnn \inch
        -R^2(x)\,d\alpha^2 - A(x)\,dz^2,
\ear
  from the set of Einstein-Maxwell equations we obtain the following equation
  for the circular metric coefficient $R(x)$:
\beq\label{krec-R(x)2}
    A^{-1}\left[\frac{R''}{R}+\frac{R'}{2R}\left(\frac{D'}{D}-
        \frac{R'}{R}\right)\right]=2\ae H_\alpha^2.
\eeq
  Here, just as in the case of a vortex gravitational field, the right-hand
  side is positive-definite, and hence $R''$ is positive where $R'=0$,
  which, as pointed out above, is a necessary condition for wormhole
  existence. The solution to the Einstein-Maxwell equations indeed describes
  a wormhole geometry:
\bearr\label{krec-reshenie_AD}
    A(x) = \frac{\ae I_z^2}{8\pi k^2}\cosh^2(kx)\cdot e^{5kx},
\nnn
    D(x) = \frac{\ae I_z^2}{8\pi k^2}\cosh^2(kx)\cdot e^{4kx},
\ear
  while the function $R(x)$, determining the length of a coordinate
  circle $x=x_0$, $z=z_0$, is obtained in the form
\beq\label{krec-reshenie_R(x)}
    R(x) = \frac{I_z}{k}\sqrt{\frac{\ae}{8\pi}}\cosh(kx)\cdot e^{kx/2}
\eeq
  (where $-\infty < x <\infty$). Here $k$ is an integration constant, $I_z$ is
  the linear axial electric current density which is a source of the
  azimuthal magnetic field. From (\ref{krec-reshenie_R(x)}) it is seen that
  $R(x)$ nowhere turns to zero, and $R(x) \to \infty$ as $x\to +\infty$ and
  as $x\to-\infty$, i.e., there is a wormhole connecting two remote regions
  of space. The wormhole throat radius is proportional to the axial electric
  current density $I_z$.

  The above results show that self-gravitating vortex fields (the spinor,
  gravitational and electromagnetic ones) can form wormholes, so that,
  for such a purpose, it is unnecessary to invoke phantom matter or scalar
  fields with negative kinetic energy, the more so as nobody knows how to
  get them. Meanwhile, as is known, a vortex azimuthal magnetic field is
  induced by a linear electric current, and a vortex gravitational field, as
  we have shown, is induced by a polarized spin of a spinor field.

  In what follows, we will show that, in addition, a rotating continuous
  medium, e.g., a perfect fluid can also be a source of a vortex
  gravitational field. To this end, we will consider the gravitational
  interaction of a perfect fluid rotating with an angular velocity
  $\omega(x^k)$ in a space-time with the metric
  (\ref{krec-tsilindr-metrica}), where, as we have shown above, there exists
  a stationary vortex gravitational field. We will show that a rotating
  self-gravitating perfect fluid can be a source of the metric
  (\ref{krec-tsilindr-metrica}). Consider the Einstein equations with
  the energy-momentum tensor of a perfect fluid
\beq\label{krec-Ain-Ideal}
    R_{ik}-\frac12Rg_{ik}=\ae\left[U_iU_k(p+\eps)-pg_{ik}\right]
\eeq
  in a space-time with the metric (\ref{krec-tsilindr-metrica}).
  Let us use the comoving reference frame, in which the 4-velocity of the
  rotating fluid has the form $U^i=(1/\sqrt{D},0,0,0)$. In this case,
  the 4-vector $U^i$ is a timelike monad vector, determining a rotating
  reference frame, and the angular velocity of the fluid is simultaneously
  the rotation angular velocity of the world-line congruence of the monad.
  In this problem setting, using the monad formalism [9], from the Einstein
  equations (\ref{krec-Ain-Ideal}) in the metric
  (\ref{krec-tsilindr-metrica}) one obtains the following equation for
  the metric coefficient $R(x)=\frac{E^2+BD}{D}$:
\bearr\label{krec-R(x)3}
    A^{-1}\left[\frac{R''}{R}+\frac{R'}{2R}\left(\frac{D'}{D}-
     \frac{R'}{R}\right)\right]=\ae(p-\eps)+\frac{4\omega^2}{c^2}.
\nnn
\ear
  It follows from this equation that if
\beq\label{krec-neravenstvo}
    \frac{4\omega^2}{c^2} > \ae(p-\eps),
\eeq
  then $R''>0$ at a point where $R'=0$, which is a necessary condition for
  wormhole existence. Therefore the inequality (\ref{krec-neravenstvo})
  is a necessary condition for wormhole formation by a rotating perfect
  fluid.

  In the case of the maximally stiff equation of state $p=\eps$, the
  condition (\ref{krec-neravenstvo}) manifestly holds. The corresponding
  exact solution to the Einstein equations (\ref{krec-Ain-Ideal}) for
  a self-gravitating rotating perfect fluid, with the equation of state
  $p = \eps$, has the following form:
\bearr\label{krec-reshen}
    A = D = 1;\qquad \omega = \omega_0 = \const;
\nnn
    p = \eps=\frac{\omega_0^2}{\ae c^2} = \const;
        \qquad
    R(x) = b^2\cosh \frac{\omega_0 x}{c};
\nnn
    -\infty< x <\infty,\qquad  b=\const.
\ear
  The solution (\ref{krec-reshen}) shows that a self-gravitating fluid
  with the maximally stiff equation of state rotates like a rigid body
  and forms a wormhole since the metric coefficient $R(x) = (BD + E^2)/D$
  of the angular part in the effective spatial metric is everywhere
  positive, and $R(x)\to\infty$ as $x \to \pm\infty$.

  That a vortex gravitational field can induce wormhole formation can be
  explained by the fact hat it has a certain energy-momentum tensor
  $T_{ik}(\omega)$ whose all components are proportional to $\omega^2$,
  and the components corresponding to pressure, $p_i(\omega)$, are negative
  [1], and it violates both the strong energy condition
  ($\eps(\omega) + p_1 + p_2 + p_3(\omega) >0$) and the weak one
  ($\eps + (p_1 + p_2 + p_3)/3 > 0$), i.e., this tensor has ``phantom''
  properties. It has the structure characteristic of a perfect fluid
  with an anisotropic negative pressure:
\bearr\label{krec-f-tenzor}
    T_{ik}(\omega) = [p(\omega) + \eps(\omega)]U_iU_k
\nnn  \cm\cm
        -(p_1 - p)\chi_i\chi_k - pg_{ik},
\ear
  where $\chi_i$ is the anisotropy vector directed along the rotation axis
  and satisfying the conditions $\chi_i U^i=0$, $\chi_i\chi^i = -1$.

  The tensor $T_{ik}(\omega)$ obeys the local conservation law:
  $T^i_k(\omega)_{;i} = 0$. The solution for the vortex intensity
  $\omega = \omega_0/(AD^{1/2})$ in \eqs (\ref{krec-Reshenie_R_ik=0}) is
  just an integral of this equation. In the case of a stationary vortex
  gravitational field in a space-time described by the metric
  (\ref{krec-tsilindr-metrica}), the energy-momentum tensor of this field
  $T_{ik}(\omega)$ has the following components:
\beq\label{krec-komponenty}
    T_k^i(\omega) = \frac{\omega^2}{\ae c^2}\cdot\diag\ (1,\ 1,\ 1,\ 3).
\eeq
  It is seen that the negative pressure  $p_z=-3\omega^2/(\ae c^2)$ along the
  rotation axis is three times as large as the radial and transversal
  pressures, $p_r = p_\alpha = -\omega^2/(\ae c^2)$, and the sum
  $\eps(\omega) + \frac 13 (p_r+p_\alpha+p_z) = -3\omega^2/(2\ae c^2) < 0$,
  i.e., the weak energy condition is violated, thus leading to possible
  wormhole existence. Besides, since the axial negative pressure is three
  times as large as the other components, at gravitational collapse of
  very massive rotating astrophysical objects (the most massive stars,
  galactic nuclei), in which process the matter density extremely grows
  along with rotation velocity, forming a vortex gravitational field like
  (\ref{krec-komponenty}), such an object will stretch along its rotation
  axis. As a result, a stable, rapidly rotating astrophysical object can
  form, having a maximally stiff equation of state and stretched along its
  rotation axis, which can be a wormhole described by the above solution
  (\ref{krec-reshen}) for a stationary rotating perfect fluid configuration.

  We can conclude that there can be a fourth final state of evolution of
  astrophysical objects (stars of various masses and galactic nuclei), in
  addition to three known states --- a white dwarf, a neutron star (pulsar),
  and a black hole. Namely, a very massive rotating object can possibly form
  a wormhole with intense rotation and an equation of state close to the
  limiting one.

  Thus we have shown that vortex fields (spinor, gravitational and magnetic)
  can form wormholes. A source of a vortex gravitational field can be
  a spinor field with polarized spin or a rapidly rotating continuous medium.
  This leads to one more possible final state of astrophysical object
  evolution, a wormhole.

  The authors thank K.A. Bronnikov for attention to the work and helpful
  comments.

}

\end{document}